\documentclass{article}
\usepackage[utf8]{inputenc}
\usepackage{hhline}
\usepackage{authblk}
\usepackage[colorlinks,linkcolor=blue,linktoc=page, citecolor=blue, urlcolor=blue]{hyperref}
\usepackage{doi}
\usepackage{graphicx}
\usepackage[numbers,sort&compress]{natbib}
\usepackage[normalem]{ulem} 
\usepackage{amsmath}
\usepackage{fancyhdr}
\usepackage{xcolor}
\usepackage{soul}
\usepackage{bm}

\title{3D simulations of vertical displacement events in tokamaks: A benchmark of M3D-C$^1$, NIMROD and JOREK}

\author[1]{\small F.J. Artola}
\author[2]{ C.R. Sovinec}
\author[3]{ S.C. Jardin}
\author[4]{ M. Hoelzl}
\author[5]{ I. Krebs}
\author[3]{ C. Clauser}

\affil[1]{ITER Organization, 13067 St. Paul Lez Durance Cedex, France}
\affil[2]{University of Wisconsin-Madison, Madison, Wisconsin 53706-1609, USA}
\affil[3]{Princeton Plasma Physics Laboratory, P.O. Box 451, Princeton, New Jersey 08543-0451, USA}
\affil[4]{ Max Planck Institute for Plasma Physics, Boltzmannstr. 2, 85748 Garching b. M., Germany}
\affil[5]{FOM Institute DIFFER - Dutch Institute for Fundamental Energy Research, De Zaale 20, 5612 AJ Eindhoven, the Netherlands}

\def\VecA{\mathbf{A}}
\def\VecV{\mathbf{v}}

\def\pres{p}

\def\VecB{\mathbf{B}}

\def\VecJ{\mathbf{J}}

\def\TensD{\underline{\pmb{\text{D}}}}
\def\TensV{\underline{\pmb{\tau}}}
\def\TensK{\underline{\pmb{\kappa}}}

\newcommand{\pderiv}[2]{\frac{\partial #1}{\partial #2}}       

\date{}

\begin{document}

\maketitle

Email: \href{mailto:javier.artola@iter.org}{javier.artola@iter.org}

\section*{Abstract}
In recent years, the nonlinear 3D magnetohydrodynamic codes JOREK, M3D-C$^1$ and NIMROD   developed the capability of modelling realistic 3D vertical displacement events (VDEs) including resistive walls. In this paper,  a comprehensive 3D VDE benchmark is presented between these state of the art codes. The simulated case is based on an experimental NSTX plasma but with a simplified rectangular wall. In spite of pronounced differences between physics models and numerical methods, the comparison shows very good agreement in  the relevant quantities used to characterize disruptions such as the 3D wall forces and energy decay. This benchmark does not only bring confidence regarding the use of the mentioned codes for disruption studies, but also shows differences with respect to the used models (e.g. reduced versus full MHD models). The simulations show important 3D features for a NSTX plasma such as the self-consistent evolution of the halo current  and the origin of the wall forces. In contrast to other reduced MHD models based on an ordering in the aspect ratio, the ansatz based JOREK reduced MHD model allows capturing the 3D dynamics even in the spherical tokamak limit considered here. 

\medskip

The following article will be submitted to Physics of Plasmas. If accepted, it will be found at \url{https://aip.scitation.org/journal/php} after it is published.

\section{Introduction}
Vertical displacement events (VDEs) are axisymmetric instabilities that arise for elongated plasmas when the control of the vertical position is lost. In future large fusion devices such as ITER, major disruptions will unavoidably lead to the loss of plasma control \cite{Gribov_2007} and a resulting VDE. Such events drive the plasma column into the wall, causing large thermal and electromagnetic loads to the plasma facing components and to the vacuum vessel \cite{Hender_2007}. During VDEs, additional 3D MHD instabilities can arise resulting in the localization of the loads and in net sideway forces. Moreover, it has been observed that these forces can rotate toroidally over time, leading to large forces on the  vacuum vessel and its supporting structures when the rotation frequency resonates with the natural frequencies of the vessel \cite{schioler2011dynamic,gerasimov2014plasma}. 

\medskip

In this respect, the development of  validated 3D MHD codes including the effect of resistive walls is crucial to study these loads for future machines. Although axisymmetric codes such as DINA \cite{khayrutdinov1993studies}, TSC \cite{jardin1986dynamic} or CarMa0NL\cite{Villone_2013}  allow studying many crucial aspects of disruptions, they cannot take into account these 3D effects. Moreover, for \textit{hot} VDEs the edge safety factor ($q_a$) is strongly reduced due to the slow decay of the plasma current during the vertical motion \cite{artola2020understanding}. In such cases, axisymmetric codes may impose $q_a \geq 1$ by forcing the current profile, otherwise $q_a$ drops below unity when the minor radius approximately drops by a factor of 2 ($q_a\sim a^2/I_p$). When $q_a$ falls below 2 or so, the configuration will no longer remain axisymmetric. Non-axisymmetric modes will grow, producing horizontal forces and stochastic field lines. The minimum $q_a$ and the level of current density is strongly influenced by MHD activity and parallel dynamics in the stochastic field line region affecting the current density directly, but also via a modification of the Spitzer's resistivity due to the fast temperature reduction.

\medskip

In recent years, simulations of 3D VDEs have been performed by Strauss \cite{strauss2018reduction} with the M3D code \cite{strauss2018reduction}, Sovinec and Bunkers \cite{sovinec2018effects} with  NIMROD \cite{sovinec2004jcp}, Pfefferlé, et al with \cite{pfefferle2018modelling} M3D-C$^1$\cite{Ferraro2016} and Artola\cite{artolasuch:tel-02012234} with JOREK\cite{huysmans2007mhd,holzl2012coupling}. However, no benchmark for 3D VDEs has been performed among these codes. Except for highly idealized cases, there are no analytical solutions for events that involve chaotic magnetic field lines and large vertical displacements. The work presented in this article is therefore an essential contribution to the verification and validation of these numerical  codes.

\medskip

In this work, we present such a benchmark for the codes JOREK, M3D-C$^1$ and NIMROD. These three codes are among the limited number of codes capable of simulating 3D MHD instabilities in tokamak geometry including resistive walls. The benchmark is  addressing the full 3D dynamics and is partly based on the axisymmetric benchmark that was already performed between these three codes \cite{krebs2020axisymmetric}. The goal of this work is to demonstrate the consistency among these codes and to provide the fusion community with a useful benchmark for MHD simulations of disruptions.

\medskip

This paper is organised as follows. In section \ref{sec:models} we revisit the models and the numerics used by the different codes. The setup and the  parameters of the benchmark case are described in section \ref{sec:setup}. The comparison of the obtained results and the analysis of the 3D VDE simulations are presented in section \ref{sec:results}. Finally, we summarize our conclusions in section \ref{sec:conclusions}.

\section{Models and codes description}
\label{sec:models}

\begin{table}[ht]

\small
\def\arraystretch{1.5}

\resizebox{\textwidth}{!}{
\begin{tabular}{||c||c|c|c|}
\hline

\textbf{Feature/Code} & \textbf{JOREK}   &  \textbf{M3D-C}$^1$ & \textbf{NIMROD} \\ \hhline{||=||=|=|=|}
\textbf{Toroidal dicretization} & Fourier harmonics & Cubic Hermite elements & Fourier harmonics \\ \hline
\textbf{$R$-$Z$ plane elements}    & \def\arraystretch{1}\begin{tabular}[c]{@{}c@{}} Quadrilateral isoparametric\\ bicubic Bezier
\end{tabular} & Triangular reduced-quintic  &    \def\arraystretch{1}\begin{tabular}[c]{@{}c@{}} Quadrilateral\\ bicubic \end{tabular}\\ \hline
\textbf{Continuity} & $G^{1*}$ & $C^1$ & $C^0$ \\ \hline
\textbf{Wall model}             & \def\arraystretch{1}\begin{tabular}[c]{@{}c@{}}Axisymmetric\\ thin wall\end{tabular}                                 &\def\arraystretch{1} \begin{tabular}[c]{@{}c@{}}Axisymmetric\\ thick wall\end{tabular}            & \def\arraystretch{1}\begin{tabular}[c]{@{}c@{}}Axisymmetric\\ thin wall\end{tabular} \\ \hline
\textbf{Vacuum region}  & \def\arraystretch{1}\begin{tabular}[c]{@{}c@{}}Un-meshed\\ Green's function method\end{tabular}  & \def\arraystretch{1}\begin{tabular}[c]{@{}c@{}}Meshed up to \\ outer boundary\end{tabular}  & \def\arraystretch{1}\begin{tabular}[c]{@{}c@{}}Meshed up to \\ outer boundary\end{tabular}   \\ \hline
\textbf{MHD model} & Reduced MHD & Full MHD & Full MHD  \\ \hline
\textbf{Time-advance} & Implicit (Gears) &  Split-implicit  & Split-implicit  \\ \hline
\end{tabular}
}

\caption{Summary of the different numerical properties of the code versions used in the present benchmark. $^*$Note that $G$ continuity stands for geometric continuity instead of parametric continuity ($C$).}
\label{tab:codes_features}
\end{table}

In this section the models used in JOREK, M3D-C$^1$ and NIMROD are discussed. In table \ref{tab:codes_features} the main properties of the codes are summarized and they are explained in more detail below.

\subsubsection*{JOREK}
JOREK is a  fully implicit non-linear extended MHD code for realistic tokamak geometries including open field-line regions~\cite{huysmans2007mhd}. JOREK discretizes the poloidal plane with quadrilateral bicubic Bezier elements using an isoparametric mapping~\cite{Czarny2008Jocp}. The variation along the toroidal direction is  represented by Fourier harmonics and  for the time advance  typically  the Crank-Nicolson  or the BDF5 Gears scheme is used. Although a full MHD model is available in JOREK \cite{haverkort2016implementation,pamela2020}, the extension for resistive walls has not been implemented yet for that physics model. The base MHD model used for this benchmark is  
\begin{align}
\pderiv{\VecA}{t} &= \VecV\times\VecB - \eta\VecJ  - \nabla \Phi, \label{eq:mhd:A}\\
\rho\pderiv{\VecV}{t} &= -\rho\VecV\cdot\nabla\VecV - \nabla p +
\VecJ\times\VecB + \nabla\cdot\TensV 
+\mathbf{S}_{\VecV},
\label{eq:mhd:v} \\
\pderiv{\rho}{t} &=
-\nabla\cdot(\rho\,\VecV)
+\nabla\cdot(\TensD\nabla\rho)
+S_\rho,\label{eq:mhd:rho}
\\
\pderiv{p}{t} &= -\VecV\cdot\nabla p
- \gamma p\nabla\cdot\VecV
+ (\gamma -1)\nabla\cdot(\TensK\nabla T) + (\gamma -1)\TensV:\nabla\VecV + S_{\pres} 
\label{eq:mhd:p}
\end{align}
with the following reduced MHD ansatz for the magnetic field ($\VecB$) and the mean plasma velocity ($\VecV$)
\begin{align}%
\VecB &= \nabla\psi\times\nabla\phi + F_0 \nabla\phi, \label{eq:B}\\
\VecV &= -\frac{R^2}{F_0}\nabla\Phi\times\nabla\phi +\VecV_\parallel, \label{eq:v}
\end{align}
where $\psi$ is the poloidal magnetic flux and $F_0 = RB_\phi$ is a constant representing the main reduced MHD assumption, which is that the toroidal field is fixed in time. Note however that poloidal currents are not fixed in time and they evolve according to the current conservation and momentum balance equations, but their contribution to the toroidal field is neglected. The calculated quantities are  the magnetic potential ($\VecA$), the ion density ($\rho$), the total pressure ($p$), the total temperature ($T$), the electrostatic potential ($\Phi$) and the current density ($\VecJ$). The other parameters are the plasma resistivity ($\eta$), the stress tensor($\TensV$), the heat and particle diffusion coefficients ($\TensK,\TensD$) and the ratio of specific heats ($\gamma$). Although they are not employed for this benchmark, the particle, momentum and heat sources $(S_\rho,\mathbf{S}_{\VecV},S_{\pres})$ can be specified in the JOREK input file. Note that the reduction of the equations is ansatz-based and does not result from geometrical ordering assumptions.

\medskip

The resistive wall and the free-boundary conditions are included by coupling JOREK to the STARWALL code \cite{merkel2015linear,holzl2012coupling, artolasuch:tel-02012234}. The coupling is performed by solving the full vacuum domain with a Green's function method and therefore the vacuum does not need to be meshed.  STARWALL uses the thin wall assumption  and discretises the wall with linear triangular elements. Although the employed wall formalism is  implemented for 3D thin walls including holes, the used  setup is restricted to an axisymmetric wall.

\subsubsection*{M3D-C$^1$}
M3D-C$^1$ is a versatile MHD code with a number of options. It can be run in the 2D (axisymmetric) mode, a linear-3D mode (for a single toroidal harmonic), or in a fully 3D nonlinear mode as was used for this paper. The magnetic vector potential and ion velocity are represented in terms of scalar quantities as $\mathbf{A}=\psi\nabla\phi + R^2\nabla\phi\times\nabla f - F_0 \ln R \nabla Z $ and $\mathbf{v}=R^2\nabla U\times\nabla \phi + w R^2 \nabla \phi + R^{-2}\nabla_\perp \chi$ . Here $(R,\phi,Z)$ are  cylindrical coordinates. The symbol $\nabla_\perp$ denotes the gradient in the $\nabla R$ and $\nabla Z$ directions. The function $f$ is constrained to be regular at the origin so that the constant $F_0$ is proportional to the total current in the toroidal field magnets. The code can be run in the 2-variable reduced MHD option where only ($\psi, U$) are advanced in time, in the 4-variable reduced MHD option where ($\psi, f, U, w$) are advanced in time, in the full single fluid MHD option (used in this paper), or in a 2-fluid MHD option.

\medskip

In the conductor region, only the magnetic vector potential variables are advanced in time, with the Ohm’s law: $\mathbf{E}=\eta_w \mathbf{J}$, where $\eta_w$ is the conductor resistivity, which may be a function of space. In the vacuum region: $\nabla\times\nabla\times\mathbf{A}=0$ is enforced.

\medskip

The code uses an unstructured mesh with triangular elements in the ($R,Z$) plane, and extruded in a structured manner in the toroidal direction. The mesh is normally adapted to the geometry and the problem, so that a finer mesh size is used in locations where large gradients are expected. All scalar variables are expanded in 3D finite elements that are a tensor product of the Bell \cite{Braess_2002} element in the ($R,Z$) plane and Hermite cubic elements in the toroidal angle $\phi$. This representation enforces continuity of each scalar and all of its first-derivatives across element boundaries.

\medskip

In the 3D nonlinear code, a split-implicit time advance is used where the velocity variables are advanced first, followed by the magnetic-field variables, the pressure(s) and then density. Implicit hyper-resistivity terms can be included in the time advance. Boundary conditions for the temperature(s) and density are set at the boundary of the plasma domain. For the magnetic field variables, the B.C.s are set  at the outermost computational boundary containing the wall and the vacuum regions.

\subsubsection*{NIMROD}
The NIMROD code solves fluid-based models for magnetized plasma in non-reduced form.  Unlike JOREK and M3D-C1, it evolves components of magnetic field and flow-velocity.  The system of equations is equivalent to Eqs.~(1-4), except that Faraday's law is used in place of Eq.~(1), the particle diffusivity in (3) is isotropic, and Eq.~(4) is replaced by a temperature equation.  Although NIMROD has capability for advancing magnetic field with 3D temperature-dependent resistivity, the NIMROD computations reported here use resistivity computed with the toroidally symmetric part of $T$.  The system is advanced in time with an implicit leapfrog method that, in comparison with fully consistent implicit methods, requires smaller but less computationally intensive steps to achieve the same level of accuracy \cite{sovinec2010jcp}.  Like JOREK, NIMROD expands $\phi$-dependent fields in toroidal Fourier harmonics.  For the poloidal plane, the expansion is 2D $C^0$ spectral elements with node spacing based on Legendre polynomials.  The underlying polynomial representation is chosen at runtime, and here all computations use polynomials of degree 3.  The representation does not satisfy the magnetic divergence constraint, identically, but the diffusive error correction term $\kappa_{divb}\nabla\nabla\cdot\VecB$ added to Faraday's law is effective at maintaining minimal error with NIMROD's spatial representation \cite{sovinec2004jcp}.  The VDE modeling uses the thin resistive wall approximation to interface with an external numerical vacuum response, where the magnetic representation is the same is in the internal plasma region.  The NIMROD computations also use the stabilization method that is described in Ref.~\cite{sovinec2016jcp}.

\subsubsection*{Boundary conditions}
Dirichlet boundary conditions are applied for the fluid variables at the plasma-wall interface. The temperature and density B.C.s are simply $\rho=\rho(t=0)$ and $T=T(t=0)$. The velocity boundary conditions are $\mathbf{v}=0$ in M3D-C$^1$ and NIMROD, but in JOREK they are $\mathbf{v}\cdot\mathbf{n}=0$, $v_\parallel=0$ and $w_\phi=0$, where $\mathbf{n}$ is the normal vector to the boundary and $w_\phi$ is the toroidal component of the vorticity.

\section{Benchmark setup}
\label{sec:setup}

\begin{figure}
  \includegraphics[width=0.99\textwidth]{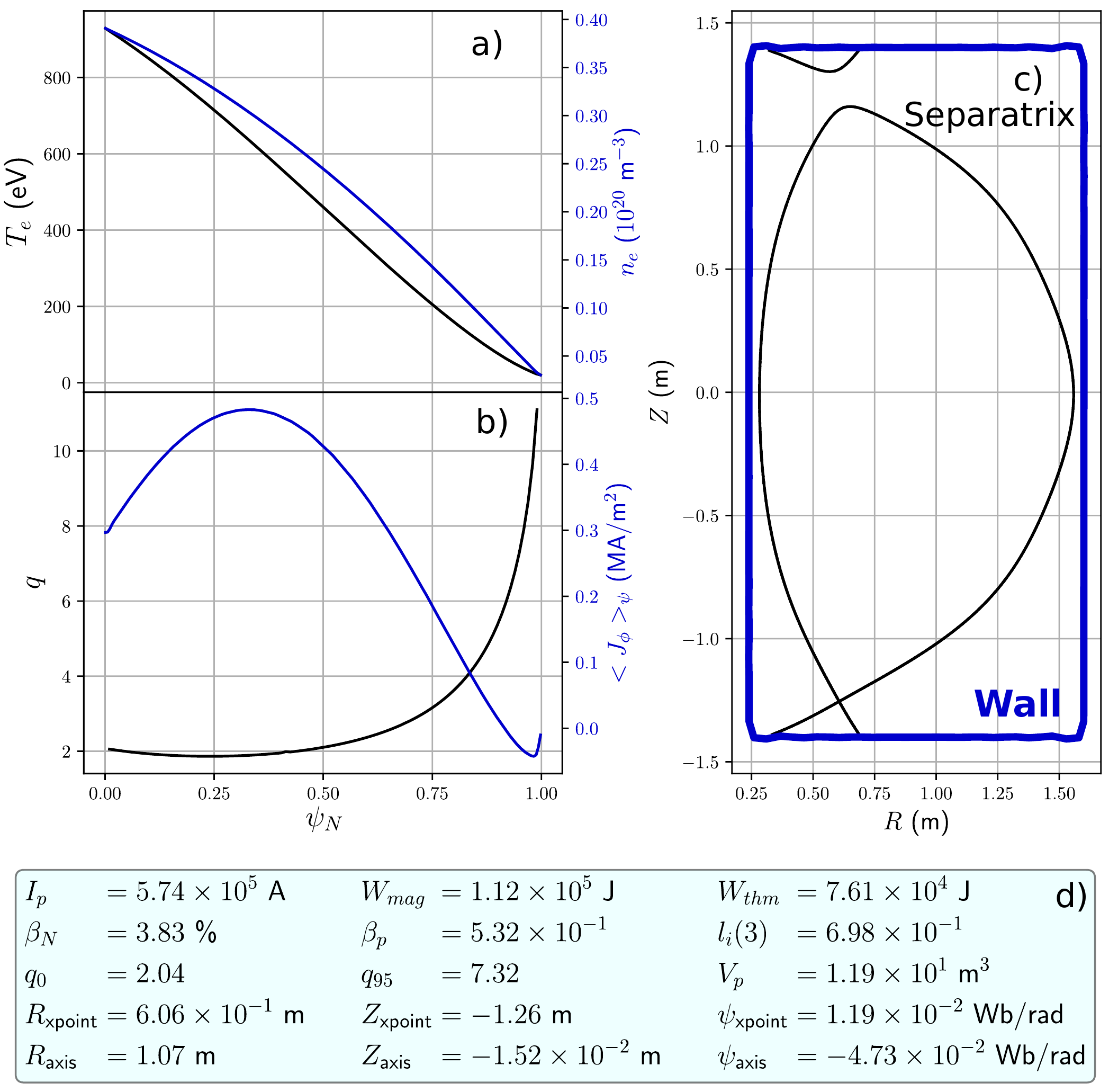}
\caption{Equilibrium used as initial condition for the benchmark. (a) Temperature and density profiles, (b) safety factor profile and poloidally averaged toroidal current density profile ($\oint_{\psi=\textrm{const}} J_\phi dl /\oint_{\psi=\textrm{const}} dl$)  as functions of the normalized poloidal flux. (c) Wall and initial separatrix geometry. (d) List of  relevant scalars describing the equilibrium.}
\label{fig:equilibrium}
\end{figure}

\paragraph{Plasma equilibrium} As mentioned in the introduction, the benchmark is a 3D version of the axisymmetric benchmark that was already published for these three codes \cite{krebs2020axisymmetric}. The chosen equilibrium is based on an NSTX experimental plasma (discharge \#139536 at $t=309$ ms) that was reconstructed with the EFIT code \cite{lao1985reconstruction}. The equilibrium profiles, the separatrix and the wall geometries and several scalar quantities describing the equilibrium are shown in figure \ref{fig:equilibrium}. The temperature profile is calculated from the pressure profile ($p$) given in the \textit{geqdsk} file with the expression $T_e = T_i = 946 \textrm{ eV } (p/p_\textrm{axis})^{0.6}$. Note that this file is available in the supplementary material [link] as well as the currents and geometry of the poloidal field coils that are needed to compute the free-boundary equilibrium. At the plasma-wall interface, the applied Dirichlet boundary conditions for the temperature and density are $T_{e,edge}=14.6$ eV and $n_{e,edge}= 2.20\times 10^{18} \textrm{ m}^{-3}$.

\paragraph{Vacuum vessel} For simplicity, a rectangular vacuum vessel is chosen instead of the complicated geometry of the NSTX vacuum vessel (see figure \ref{fig:equilibrium} (c)). The rectangular shape is defined by the four vertices $(R = 0.24 \textrm{ m}, Z = \pm 1.4 \textrm{ m})$  and $(R = 1.6 \textrm{ m}, Z = \pm 1.4 \textrm{ m})$. Note that the wall corners are not exactly sharp in the simulations because Fourier harmonics are used to represent the wall contour. The  thickness of the resistive wall is $\Delta_w = 0.015$ m and the wall resistivity is $\eta_w = 3\times 10^{-5} \Omega $ m. Where a thin-wall approximation is used, the effective thin wall resistivity $\eta_w/\Delta_w$ is used.

\begin{table}[ht]
\caption{Physical parameters used for the benchmark case.}

\vspace{0.3cm}

\begin{tabular}{|ll|}
\hline
\multicolumn{2}{|c|}{\textbf{During all phases} } \\ \hline
No loop voltage & \\
No Ohmic heating & \\
No radiation & \\
No heating and particle sources & \\
Ion mass:      &$m_i = 2 m_\textrm{proton}$    \\
Ion charge:    &$Z = 1$    \\
Ion-electron temperature ratio:    &$T_e / T_i = 1$    \\
Viscosity: &$\nu_\parallel = \nu_\perp = 5.16 \times 10^{-7}$ kg (ms)$^{-1}\textrm{*}$ \\
Resistivity: &$\eta_\parallel = \eta_\perp = 1.75\times 10^{-2} T_e[\textrm{eV}]^{-3/2} \Omega$m \\  \hline
\end{tabular}

*In the NIMROD computations, $\nu_\parallel$ is increased to $100 \times \nu_\perp$ at 1.08 ms into the 3D phase to avoid overshoot in the $n$-expansion.

\vspace{0.3cm}

\begin{tabular}{|ll|}
\hline
\multicolumn{2}{|c|}{\textbf{During the 2D phase} (before the plasma becomes limited) } \\ \hline
Heat diffusion coefficients:      &$\kappa_\perp = 10^{-5}\kappa_\parallel = 1.54 \times 10^{19}$ (ms)$^{-1}$   \\
Particle diffusion coeffcients:    &$D_\perp =D_\parallel=1.54 \textrm{ m}^2/\textrm{s}$  \\  \hline
\end{tabular}

\vspace{0.3cm}

\begin{tabular}{|ll|}
\hline
\multicolumn{2}{|c|}{\textbf{During the 3D phase} (after the plasma becomes limited) } \\ \hline
Heat diffusion coefficients:      &$\kappa_\perp = 10^{-5}\kappa_\parallel = 2.35 \times 10^{21}$ (ms)$^{-1}$     \\
Particle diffusion coeffcients:    &$D_\perp= D_\parallel = 40  \textrm{ m}^2/\textrm{s}$  \\  \hline
\end{tabular}
\label{tab:params}
\end{table}

\paragraph{Simulation phases and parameter choice} The simulation is divided into two phases: an axisymmetric (2D) run and a 3D run. The values for different parameters common to both phases and specific to each phase are specified in table \ref{tab:params}. In the axisymmetric phase, the X-point drifts  downwards until the tokamak becomes limited by the lower part of the wall. The end of this phase and the start of the 3D phase is therefore determined by the latter change of geometry of the LCFS. The 2D phase is identical to the initial phase for the non-linear benchmark presented in \cite{krebs2020axisymmetric} but with the wall resistivity, plasma resistivity and particle and heat diffusion coefficients increased by a factor of 10.  Note that during the full simulation the plasma resistivity is a factor 10 larger than the Spitzer's value. Although the main goal of this work is not an experimental comparison, the chosen parameters lead to vertical displacements of $1$ m in a time-scale of $10$ ms, which is similar to the VDEs observed in experiments \cite{Gerhardt_2013}  where such displacements  take place within 5-20 ms. In the 2D phase, the heat and particle diffusion coefficients are such that the thermal energy does not decay during the vertical motion. As the parallel density transport is typically governed by parallel convection, for simplicity the parallel diffusion coefficient is set to be equal to the perpendicular particle diffusion coefficient. In the 3D phase, the coefficients are significantly increased to smooth the sharp pressure gradients that arise due to the fast shrinking of the LCFS. Although for that phase the diffusion coefficients do affect the thermal energy decay, it will be shown in the next section that the final collapse of plasma energy is governed by the 3D MHD activity once the magnetic field topology becomes chaotic.

\begin{figure}
\centering
  \includegraphics[width=1.0\textwidth]{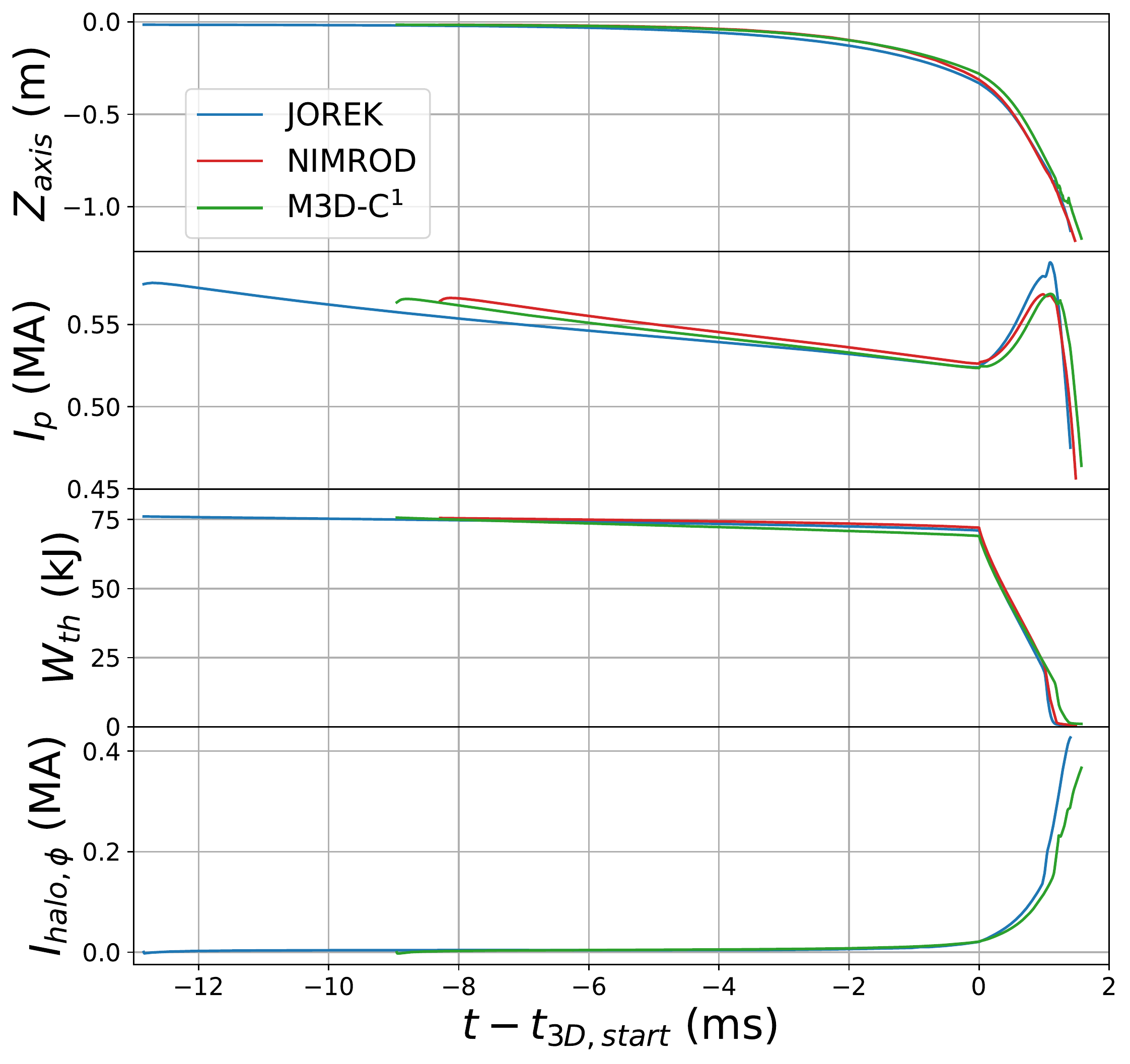}
\caption{Vertical position of the magnetic axis, plasma current and thermal energy (for all plasma domain) and toroidal halo current as function of time for the three codes. NIMROD's post-processing tools regarding computations that separate the open and the closed field line regions (e.g. $I_{halo,\phi}$) are specially challenging due to the magnetic field representation and are not yet developed for VDEs.}
\label{fig:ip_zaxis}
\end{figure}

\subsubsection*{Numerical resolution}
In JOREK, a polar grid is used with increased resolution in the region where the plasma becomes limited by the wall. The number of Bézier elements used in the plasma region is 22000, and the number of linear triangular elements to mesh the thin wall is 48000. For the toroidal direction, 11 Fourier modes were used with  $n\in [0,10]$. A resolution scan in the number of Fourier modes is shown in the next section. Time-steps of the order of 5-10 Alfvén times were used during the 2D phase and time steps of 0.1-0.2 Alfvén times were used for the 3D phase.

\medskip

The M3D-C$^1$ calculation used the unstructured poloidal plane mesh shown in \cite{krebs2020axisymmetric}, which has 17424 vertices on each plane, each with 12 degrees of freedom for each scalar field in 3D (6 in 2D). The 3D phase used 16 toroidal Hermite cubic finite elements for each of the triangular vertices. A time step of $\Delta t=0.5 \tau_A$ was used, except for a period of 0.048 ms, starting at time 10.185 ms (1.233 after the start of the 3D) when it was halved to  $\Delta t=0.25 \tau_A$, to avoid numerical instability. After this time, a “upwind” second order toroidal diffusion term was added to the scalar convection terms for the pressure, density, and magnetic field, equal to $ 0.125 \times \delta x \times v_\phi$, where $\delta x = 0.4$ m was the approximate distance between toroidal planes.

\medskip
The NIMROD computations have been run with 39000 quadrilateral bicubic elements for the region inside the resistive wall.  The meshing is largely rectangular, but a packed layer of smaller elements is used near the resistive wall and conforms to the curved corners.  The resistive wall is treated with the thin-wall model that is described in Ref.~\cite{sovinec2018effects}, and the outer vacuum region is meshed and shaped to closely match M3D-C$^1$ computations \cite{krebs2020axisymmetric}.  The 3D phase of the computations has been completed with toroidal harmonics $n\in [0,10]$, and the most dynamic interval is recomputed with $n\in [0, 21]$.  Although NIMROD's implicit leapfrog advance uses implicit advection, the temporal staggering tends to lose accuracy for dynamics if the timestep does not satisfy the Courant-Friedrichs-Lewy (CFL) condition for flow-velocity.  Thus, the timestep is adjusted, dynamically, to satisfy this condition.  For the most dynamic part of the 3D computations $\Delta t>0.015\tau_A$, which is hundreds of times larger than an explicit stability limit.  In other parts of the 3D evolution, timesteps are more than an order of magnitude larger.

\section{Results and analysis}
\label{sec:results}
In this section, the simulated case is presented, the results are compared between the three codes and the physics of the case is analysed. In subsection \ref{sec:phases_results}, the results for the different phases of the simulation are shown together with a discussion on MHD stability. In subsection \ref{sec:resolution_scan}, a resolution scan is performed to demonstrate that the case is sufficiently converged in toroidal resolution and in subsection \ref{sec:additional_physics}, the origin of the wall forces is explored as well as the influence of the parallel heat conductivity and viscosity on the results.

\subsection{Simulation phases and code comparison}
\label{sec:phases_results}

\subsubsection{Evolution overview}
We compare the vertical position of the magnetic axis, the total toroidal current (including the halo current) and the total thermal energy for the three codes in figure \ref{fig:ip_zaxis}. The total toroidal halo current is also shown for JOREK and M3D-C$^1$. As expected for the 2D phase ($t'\equiv t-t_\textrm{3D,start}<0$), the level of agreement is similar to  the axisymmetric VDE benchmark reported in \cite{krebs2020axisymmetric}. The time axis has been shifted so $t'=0$ when the plasma becomes limited in the 3 codes. As it was shown in \cite{krebs2020axisymmetric}, such  synchronization is necessary because the different numerical  perturbations initiating the VDE can lead to different time scales even if the growth rates are in good agreement.

\begin{figure}
\centering
  \includegraphics[width=1.0\textwidth]{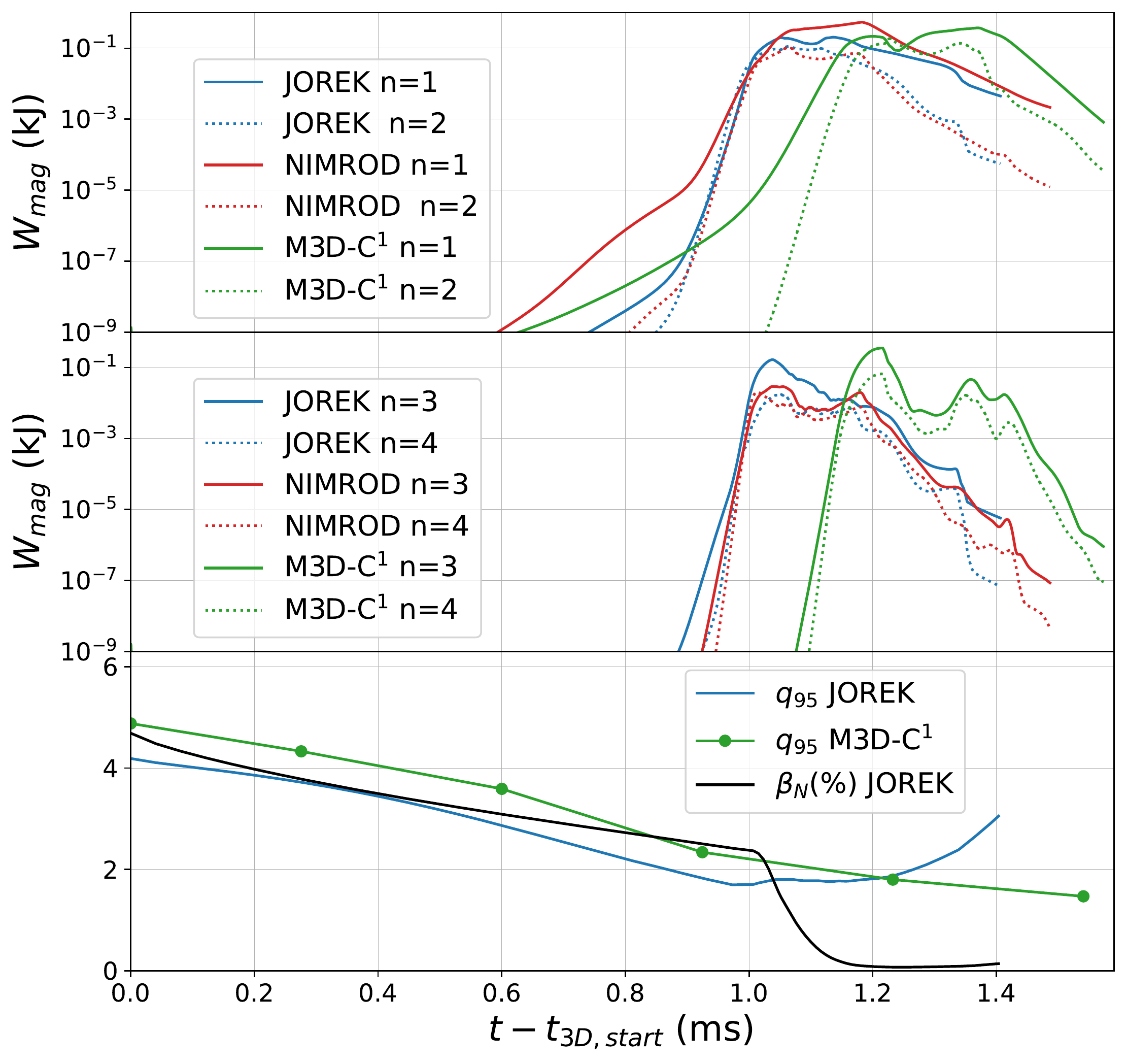}
\caption{Magnetic energy of the toroidal harmonics  $n\in [1,4]$ in the three codes as a function of time. The time traces of the edge safety factor $q_{95}$ (JOREK and M3D-C$^1$) and the normalized $\beta_N$ (JOREK) are also shown. NIMROD's post-processing tools regarding flux-surface computations (e.g. $q_{95}$) are specially challenging due to the magnetic field representation and are not yet developed for VDEs.  }
\label{fig:energies_q}
\end{figure}

In the presented case, the complete thermal quench is induced by the vertical displacement (see next section) and takes place when the plasma volume has decreased by more than a factor of two ($V_p/V_{p0} = 0.26$ at $t'=1.16$ ms in JOREK). Due to this sequence of events, the case can be classified as a \textit{hot} VDE, which features the largest wall forces observed in experiments. During the 2D phase, the plasma current and the thermal energy do not decay due to the high plasma temperature and the small diffusion coefficients. The increase of the diffusion coefficients at the start of the 3D phase leads to a loss of a large fraction of the thermal energy. Note, however, that once 3D MHD instabilities set in, the decay rate of the thermal energy is dominated by the 3D effects and not by the choice of diffusion coefficients (see $\beta_N$ time trace in figure \ref{fig:energies_q}). Such an increase in the diffusion coefficients was necessary to prevent the formation of large edge pressure gradients that arise when the plasma volume decreases. Simulations without the increase in the diffusion coefficients showed that high-$n$ edge localized modes become unstable due to the mentioned gradients. To account for the high toroidal mode numbers adequately, the numerical resolution should be largely increased and additional terms (e.g. diamagnetic flows terms) should be included in the equations augmenting the complexity of the benchmark. As this work is a benchmark exercise which particularly aims to compare the 3D wall forces (which  are observed to be given by low-$n$ mode numbers), the high-$n$ edge localized modes are avoided with the choice of the diffusion coefficients.  

\medskip
\subsubsection{MHD stability}
Linear stability calculations were performed at different vertical positions during the 2D phase. All three codes find that the initial equilibrium is unstable to resistive edge instabilities localized at the $q=3$ and $q=4$ rational surfaces, but as the equilibrium profiles evolve due to current diffusion and the vertical motion, these modes are stabilized. For example, when performing the stability calculations at $Z_\textrm{axis}=-0.15$ m, corresponding to $t'=-1.4$ ms in figure  \ref{fig:energies_q}, no instability was found.

\begin{figure}
  \includegraphics[width=1.\textwidth]{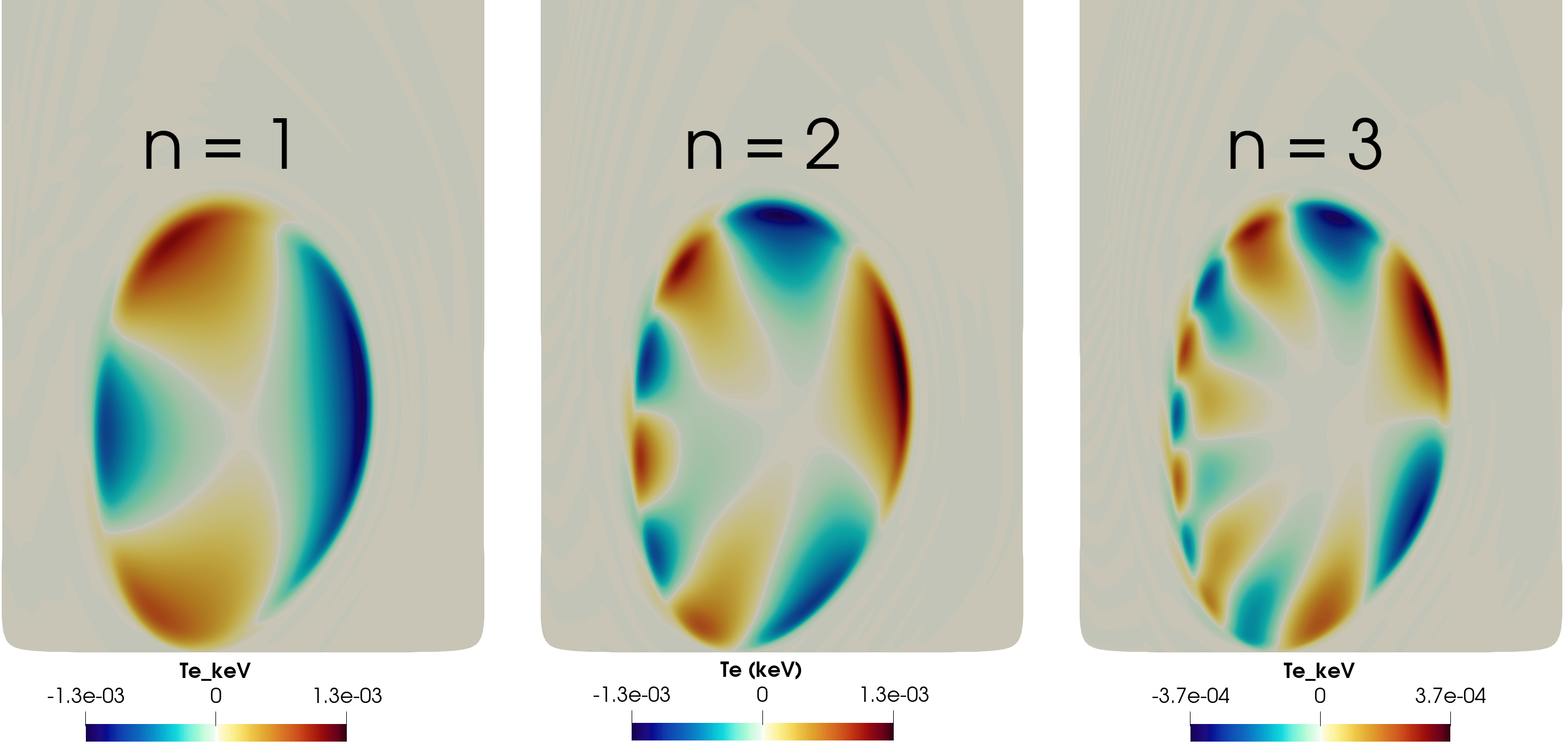}
\caption{$T_e$ mode structure for the toroidal harmonics with the largest amplitude at $t-t_{3D,start}=0.932$ ms in JOREK. }
\label{fig:mode_structure}
\end{figure}

\begin{figure}
  \includegraphics[width=0.9\textwidth]{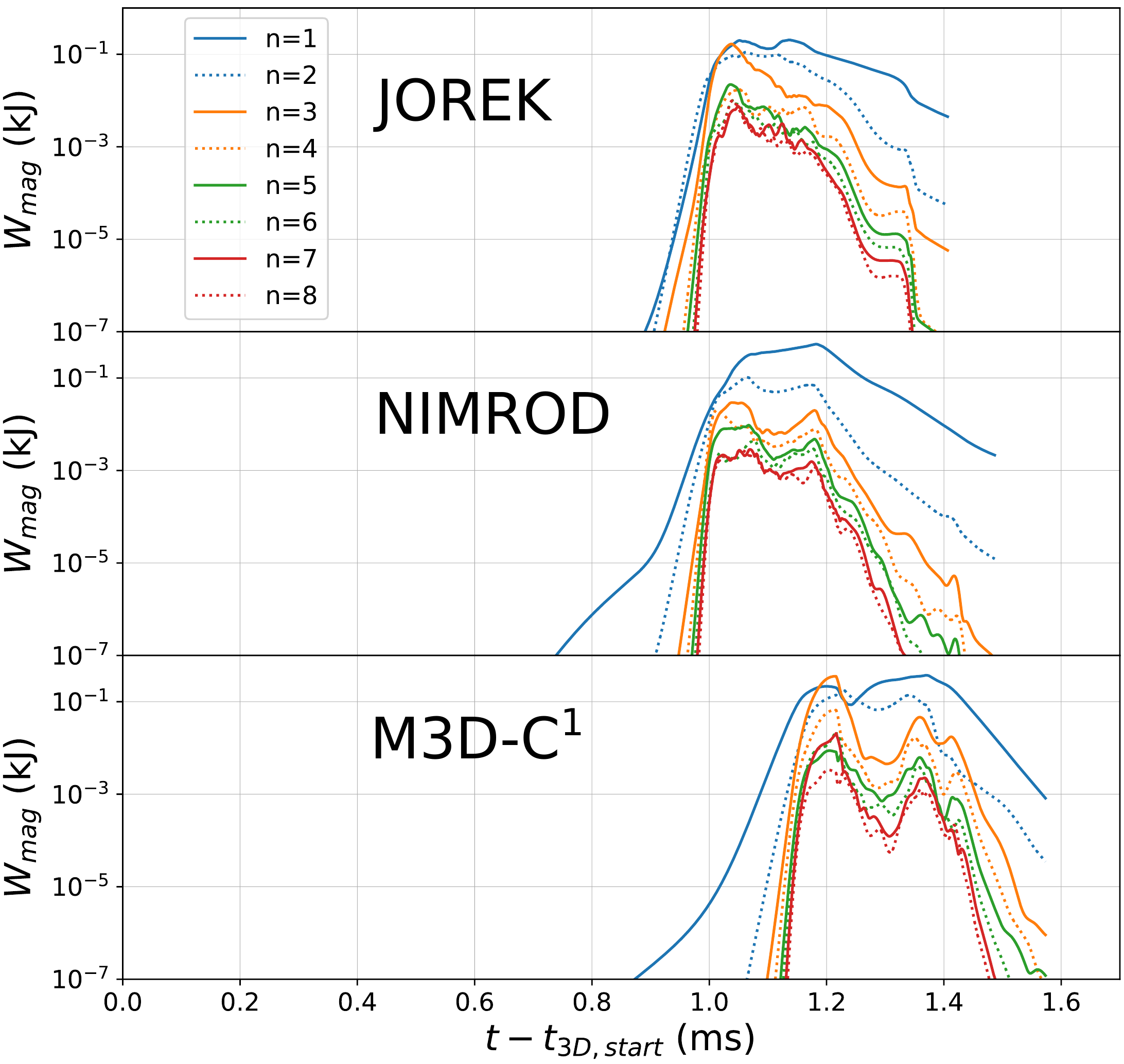}
\caption{Magnetic energies from the modes $n\in [1,8]$ in JOREK, NIMROD and M3D-C$^1$ over time.}
\label{fig:energies_3codes}
\end{figure}

In figure \ref{fig:energies_q}, the magnetic energies of the dominant toroidal harmonics  is shown as a function of time. The magnetic energy of a toroidal harmonic $n$ is defined as $W_{mag}^{n}\equiv (\int |\mathbf{B}_{pol}^n|^2 dV) / (2\mu_0)$, where $\mathbf{B}_{pol}^n$ is the poloidal magnetic field contribution given by the harmonic $n$. Figure \ref{fig:energies_q} shows that when $q_{95}$  has dropped to a value around 2, several toroidal harmonics become unstable.  Before this time, a weakly growing $n=1$ mode is observed, which is a mix of an external $m/n=3/1$ kink mode and a resistive 2/1 mode. Once the $q=2$ surface moves into the open field line region, low-n external kink modes become unstable. In figure \ref{fig:mode_structure}, the mode structures of $n=1,2,3$  are shown and it can be seen that these instabilities are associated with the $q=2$ surface  since the 2/1, 4/2 and 6/3 structures are dominant. Note that this is consistent with external kink modes \cite{freidberg2014ideal} that become unstable when 

\begin{equation}
    0< n\,q_a < m
\end{equation}
where $q_a$ denotes the cylindrical edge safety factor; in our case $q_a\approx1.8< m/n=2$. Higher-n modes  are excited by non-linear coupling,  however higher $n$ modes remain sub-dominant (see figure \ref{fig:energies_3codes}).

\medskip

All three codes show that the plasma becomes unstable  between 0.85-1.10 ms after the plasma becomes limited (see figures \ref{fig:energies_q} and \ref{fig:energies_3codes}). Also there is an agreement on the fact $n=1$ is the  dominant mode throughout the  whole dynamics. Note that at the beginning of the saturation phase, the  $n=2$ mode is also important and moreover the $n=3$ mode can exceed the $n=1$ energy in JOREK and in M3D-C$^1$ during short transient phases. 

\begin{figure}
\centering
\includegraphics[width=0.7\textwidth]{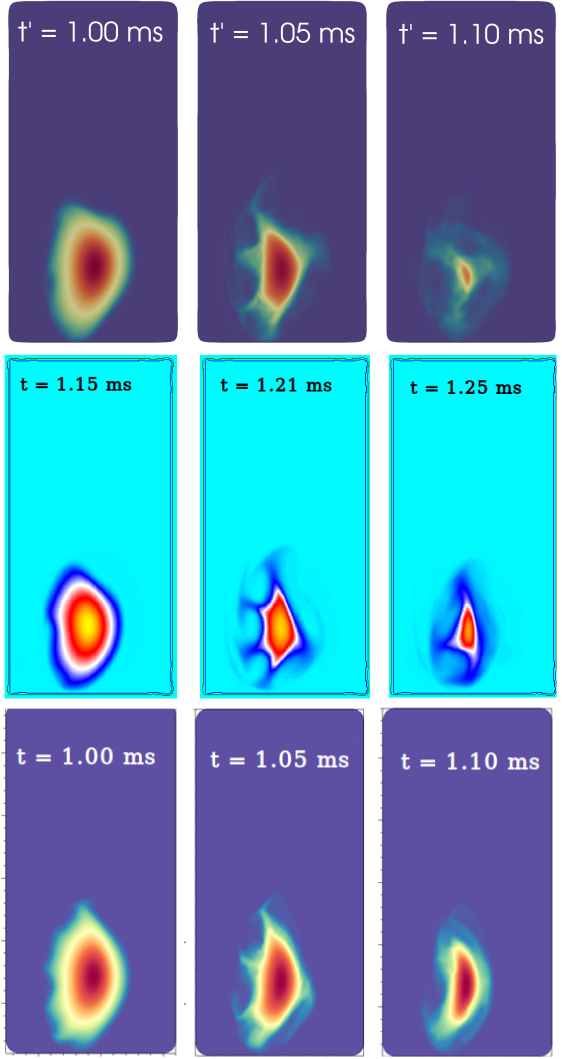}
\caption{Evolution of the pressure at the plane $\phi=0$ in JOREK (top), M3D-C$^1$ (middle) and NIMROD (bottom) in arbitrary units. }
\label{fig:pres_evol}
\end{figure}

\begin{figure}
\includegraphics[angle=0,width=1.0\textwidth]{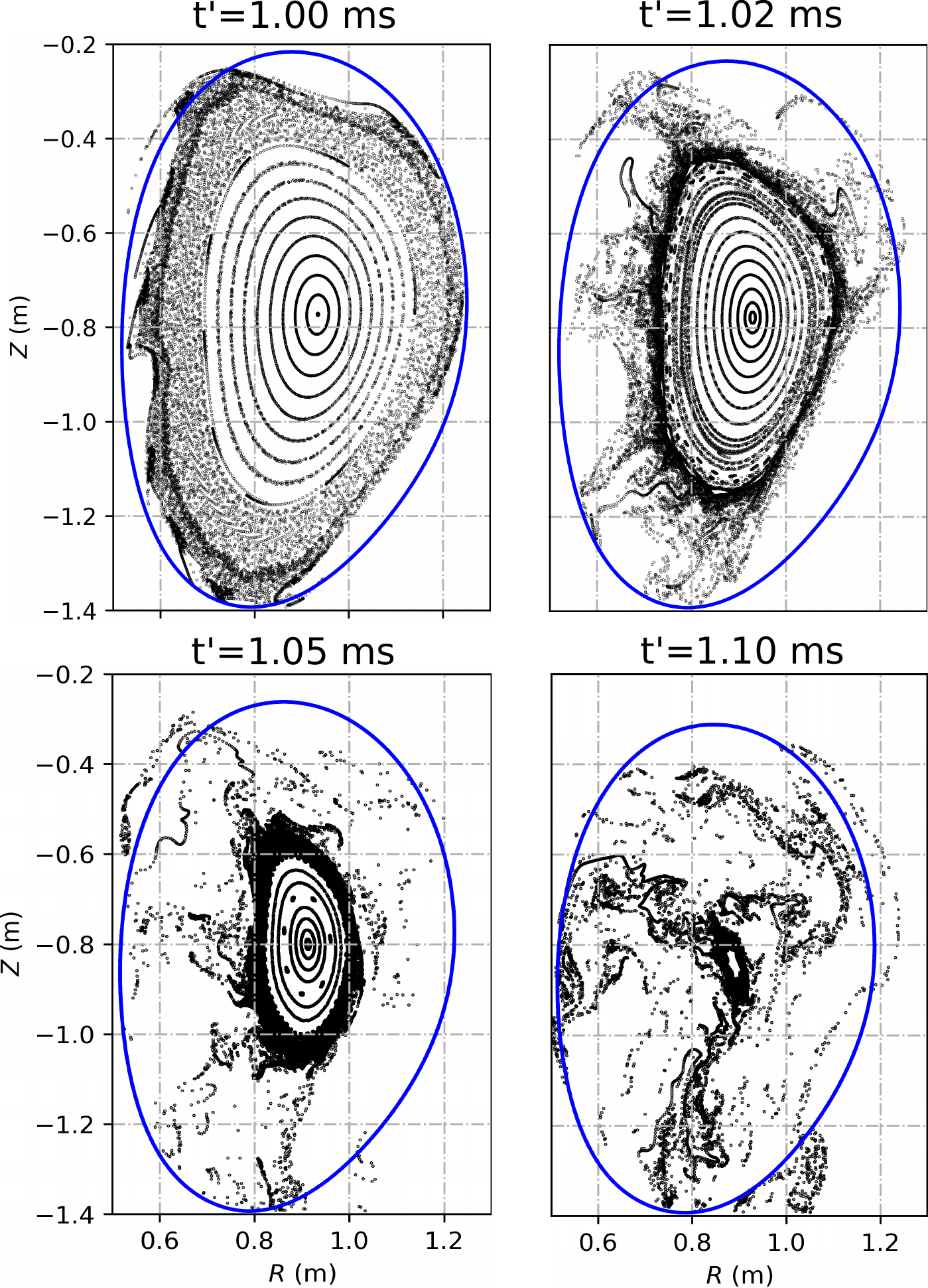}
\caption{Poincare plots for different times in the JOREK simulation. The $n=0$ LCFS is given by the blue line for reference. }
\label{fig:poincare}
\end{figure}
\subsubsection{Thermal quench}
 The evolution of the thermal pressure during the thermal quench is shown in figure \ref{fig:pres_evol}. The pressure evolution indicates that the external kink modes grow from the plasma edge and start penetrating into the plasma core. The latter process can be observed  in the Poincaré plots shown in figure \ref{fig:poincare}. The external modes deform the plasma boundary and create an ergodic layer from about $t'=1.00$ ms onwards. As these modes grow towards the plasma core, several core plasma regions become areas with open field lines.  This can be observed in the empty zones of the Poincaré plots, which imply that the field lines intersect the wall before completing a full toroidal turn. The finger like structures show regions with high density of magnetic field lines, which can do several toroidal turns before intersecting the wall. These structures can also be observed in the pressure plots of figure \ref{fig:pres_evol}. The regions with open field lines are the ones losing pressure at a faster rate and therefore they have a lower pressure (figure \ref{fig:pres_evol}). This is because the thermal energy is quickly lost along the magnetic field lines due to the fast parallel heat conduction. The fast loss of thermal energy is sensitive to the magnetic topology evolution, and in the simulations, it takes 0.14 ms in JOREK, 0.24 ms in M3D-C$^1$  and 0.20 ms in NIMROD. This range of times is consistent with NSTX disruptions where the loss of core confinement occurs in a time-scale of $\sim 0.2$ ms \cite{Gerhardt_2013_dis}.


\begin{figure}
  \includegraphics[width=0.9\textwidth]{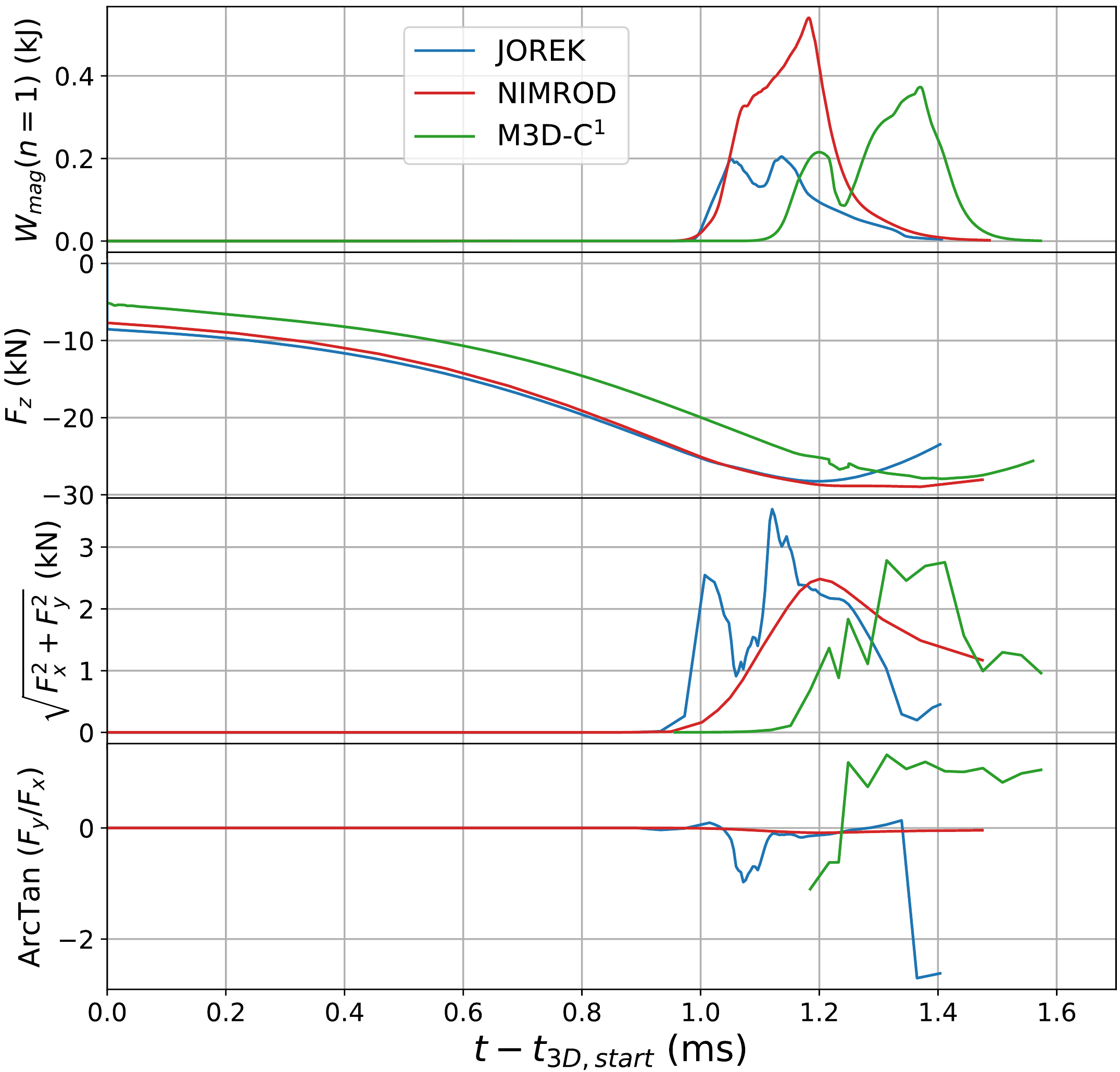}
\caption{Time traces of the $n=1$ magnetic energy, the vertical force, the total horizontal force and the toroidal phase of the total horizontal force.}
\label{fig:forces}
\end{figure}

\subsubsection{Wall forces}

In JOREK and NIMROD the total wall force is calculated with the expression given by \cite{Pustovitov_2017}

\begin{equation}
    \mathbf{F} \equiv \int_{wall} \mathbf{J}\times\mathbf{B} dV = \frac{1}{\mu_0} \oint_{wall+} \left(  (\mathbf{B}\cdot\mathbf{n})\mathbf{B} - \frac{B^2}{2}\mathbf{n}   \right) dS
\end{equation}
where $\mathbf{n}$ is the normal vector to a closed toroidal surface enclosing the wall. The force in M3D-C$^1$ is calculated directly with the integral over the vessel elements ($\int  \mathbf{J}\times\mathbf{B} dV$). In figure \ref{fig:forces}, the total vertical force, the total horizonal force and the toroidal phase of the horizontal force in the three codes are shown as a function of time (including the magnetic energy of the $n=1$ mode for reference). The vertical force ($F_z$) is already significant before the onset of the instabilities, indicating that it is dominated by 2D effects. The magnitudes and the time evolution of this force are in good agreement between the codes. Similarly the horizontal force $F_h = \sqrt{F_x^2 + F_y^2}$ shows a good agreement in magnitude. From the JOREK simulation, it can be observed that $F_h$ is correlated with the $n=1$ magnetic energy. The evolution of the $n=1$ amplitude shows two peaks that are correlated with the two peaks in the horizontal force. The toroidal phase of the horizontal force indicates that $F_h$ is not able to complete a full toroidal turn in any of the codes after the thermal quench, showing that in this case the force is rotating only very slowly. In NIMROD, no rotation is observed at all while in JOREK and M3D-C$^1$, some rotation is observed in between the peaks of the $n=1$ energy.

\begin{figure}
\centering
  \includegraphics[width=0.8\textwidth]{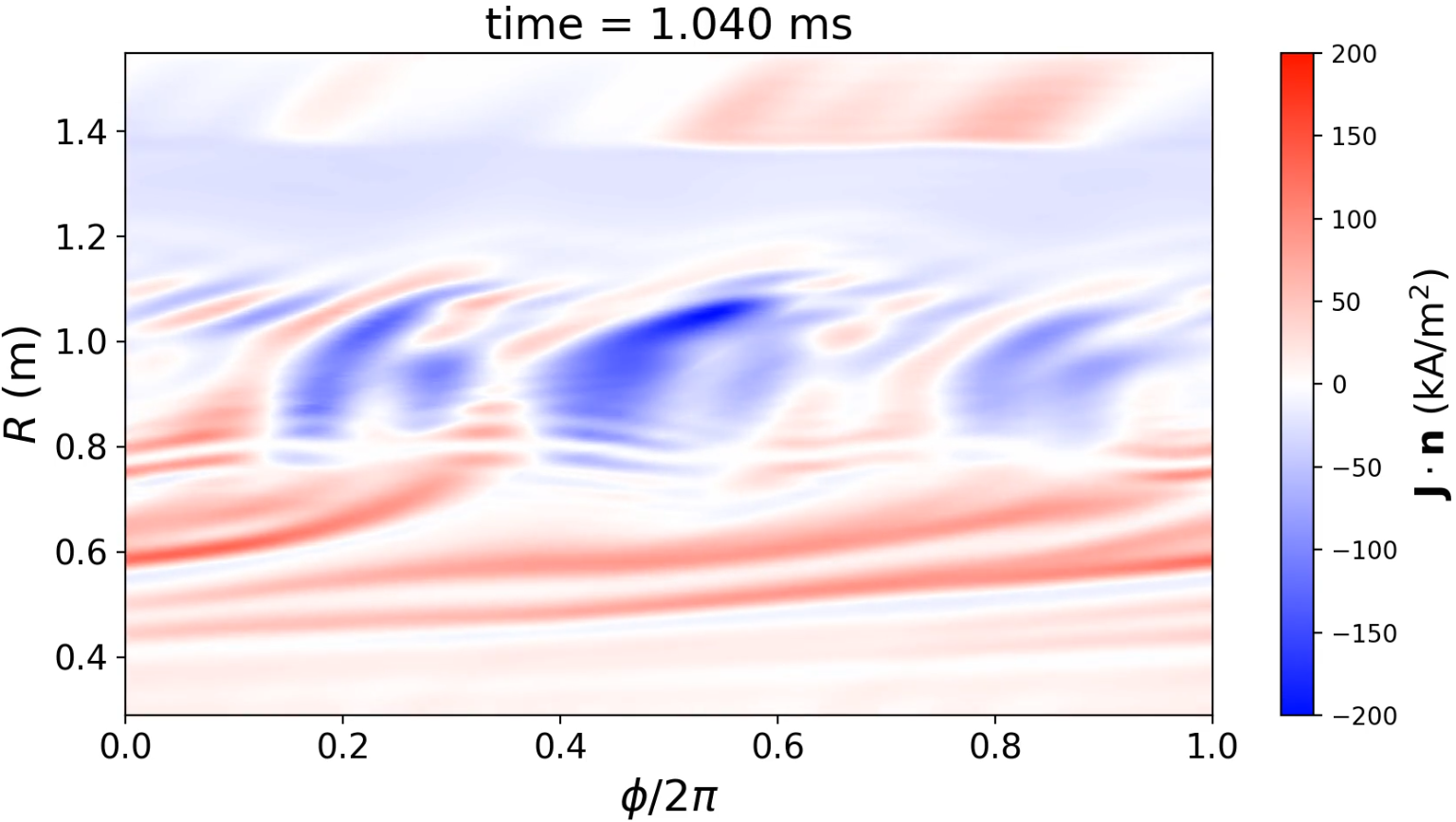}\\
  \includegraphics[width=0.8\textwidth]{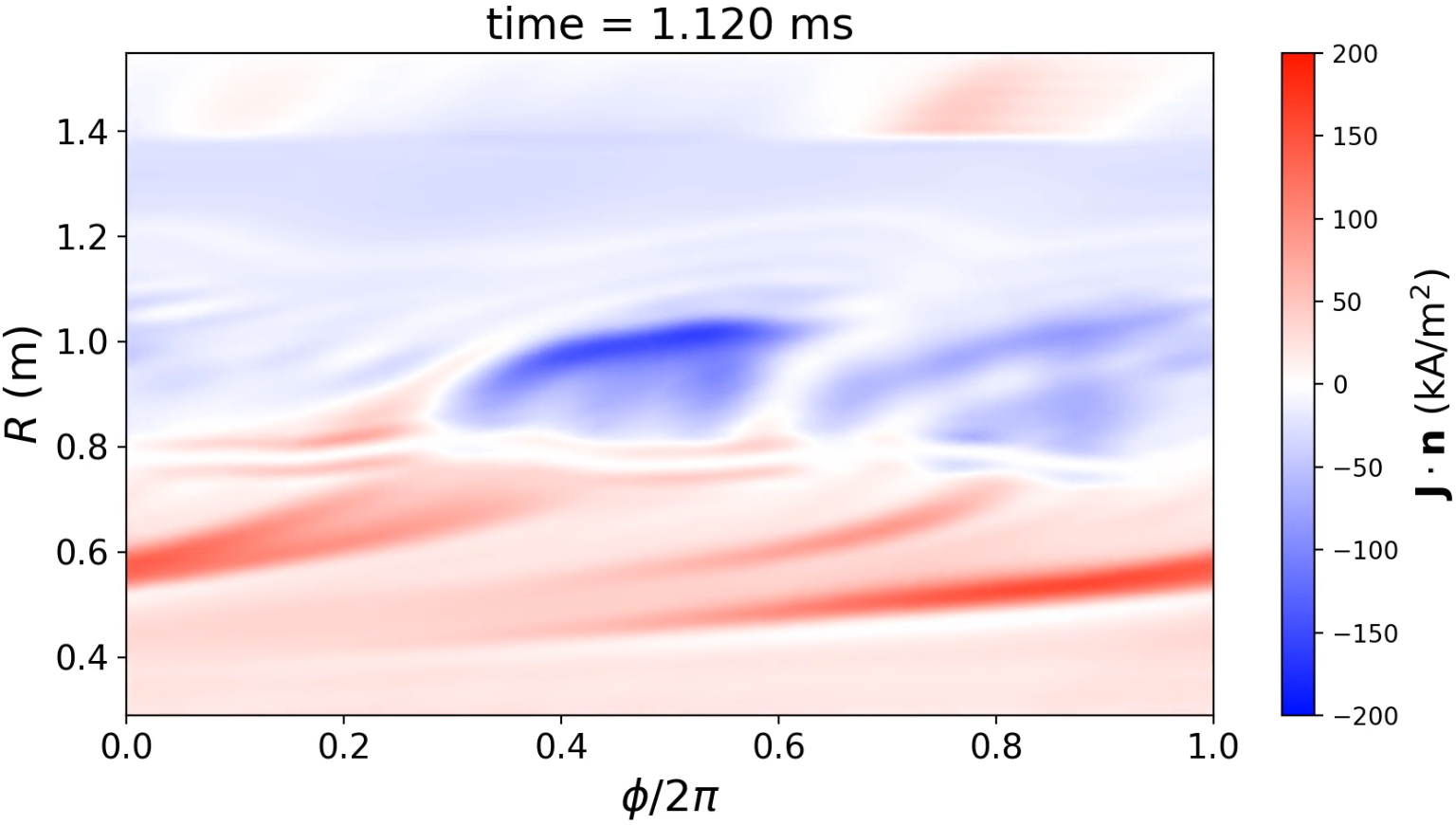}
\caption{Normal current density at the bottom side of the rectangular wall ($Z=-1.4$ m) as function of the toroidal angle $\phi$ for two times during the thermal quench in JOREK. }
\label{fig:Jnorm}
\end{figure}

\subsubsection{3D halo currents}
One of the main features that these codes offer is the capability to calculate self-consistently 3D halo currents. In figure \ref{fig:Jnorm}, the normal current density to the wall is shown as a function of the toroidal angle ($\phi$) and the $R$ coordinate on the bottom side of the rectangular wall ($Z=-1.4$ m). At time $t=1.04$ ms filamentary structures appear implying that the $n>1$ harmonics are important to determine the 3D halo current. In particular, a strong $n=3$ mode structure is present in the normal current density as this mode has the largest energy at this time in JOREK (see figure \ref{fig:energies_3codes}).  Note that across the limiter point ($R\sim 0.8$ m), the current filaments are strong enough to change the sign of the $n=0$ current density. Later, when the horizontal force reaches a maximum ($t=1.12$ ms), the halo currents show a dominant $n=1$ structure.

\begin{figure}
\centering
  \includegraphics[width=0.75\textwidth]{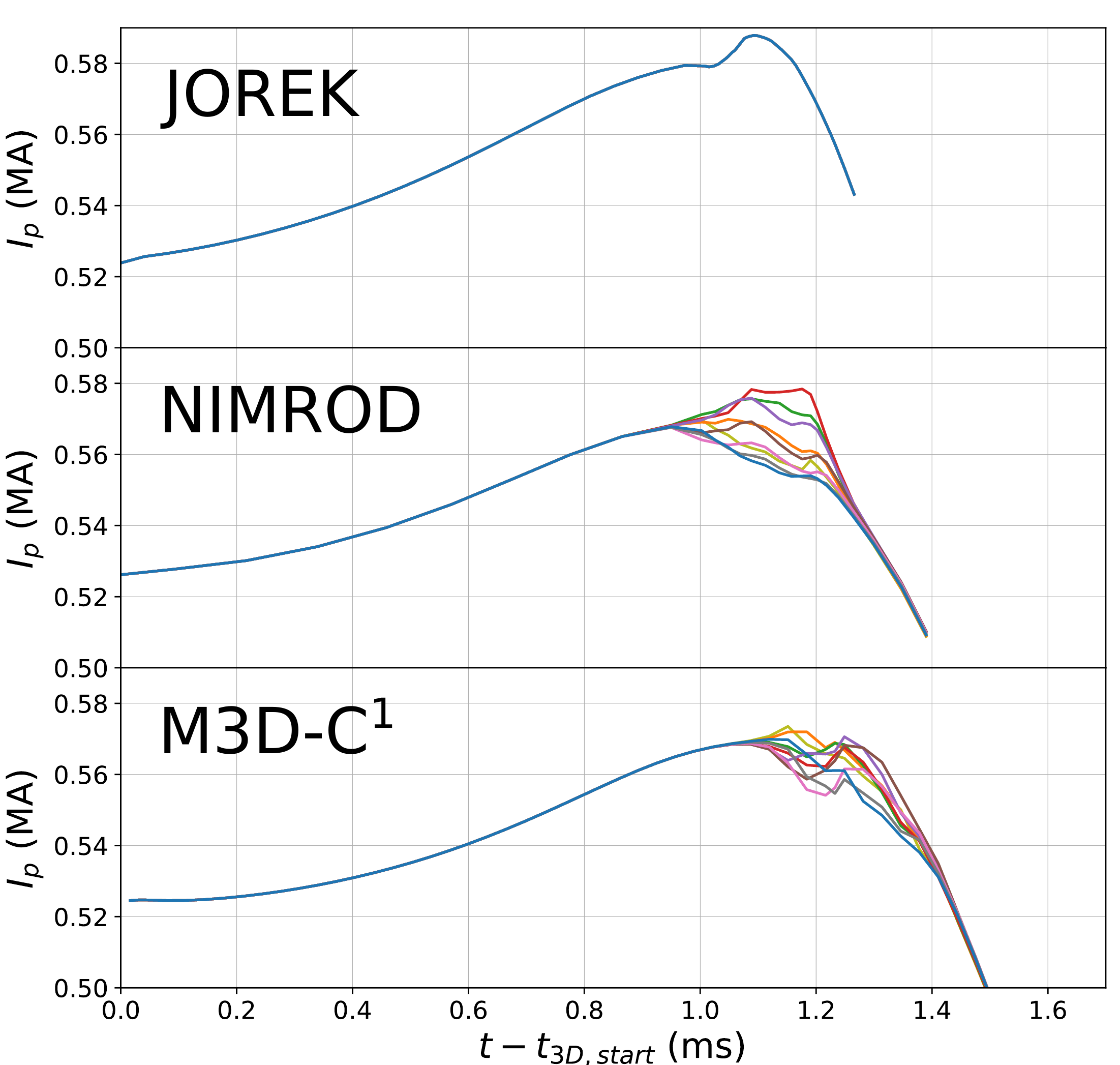}
\caption{Total toroidal current ($I_p$) computed at 10 equidistiant poloidal planes with $\phi=2\pi k/10$ and $k=0,1..9$ in JOREK, NIMROD and M3D-C$^1$. }
\label{fig:Ip_asym}
\end{figure}

\subsubsection{Toroidal plasma current asymmetries}
\label{Ip_asym}
Asymmetric VDEs are often characterized by the toroidal asymmetries of the plasma current ($I_p$). Experimental evidence shows that when measuring $I_p$ at different toroidal locations \cite{gerasimov2015jet} (e.g. using Rogowski coils at different toroidal sectors), VDEs can cause asymmetries on the plasma current of the order of 10-20\% ($\Delta I_p/I_p$). In figure \ref{fig:Ip_asym}, the plasma current is computed at 10 equidistant poloidal planes ($\phi=2\pi k/10$ with $k=0,1..9$) in the three codes. In NIMROD, the maximum $I_p$ asymmetries are of the order of 3-4\%, in M3D-C$^1$ they are on the order of 1-1.4\%, and in JOREK they are smaller than 0.01\%. 

\medskip

Although it does not seem to affect key quantities such as the wall forces, the fact that JOREK shows no $I_p$ asymmetries has been investigated  as part of this work. The poloidal electric field in the reduced MHD model is given by $\mathbf{E}_\textrm{pol}=-\nabla_\textrm{pol} \Phi$ and the boundary condition for the normal velocity implies that on the boundary $\Phi = \textrm{const}$. Therefore, the JOREK's boundary behaves as an ideal conductor in the poloidal direction. Although the normal current to the boundary can be calculated and it is well defined (see figure \ref{fig:Jnorm}), the $\Phi = \textrm{const}$ boundary condition implies that poloidal surface currents exist at the boundary even if they do not appear explicitly. In other words, no current is leaving or entering the JOREK's plasma domain, which leads to the fact that the total current must be conserved at each poloidal plane within the plasma ($\partial_\phi I_p=0$). In NIMROD and M3D-C$^1$, the poloidal electric field is finite $\mathbf{E}_\textrm{pol}=\eta \mathbf{J}_\textrm{pol}$ and therefore the normal current can leave or enter the plasma domain.

\subsection{Resolution scans}
\label{sec:resolution_scan}
\begin{figure}
\centering
  \includegraphics[width=0.8\textwidth]{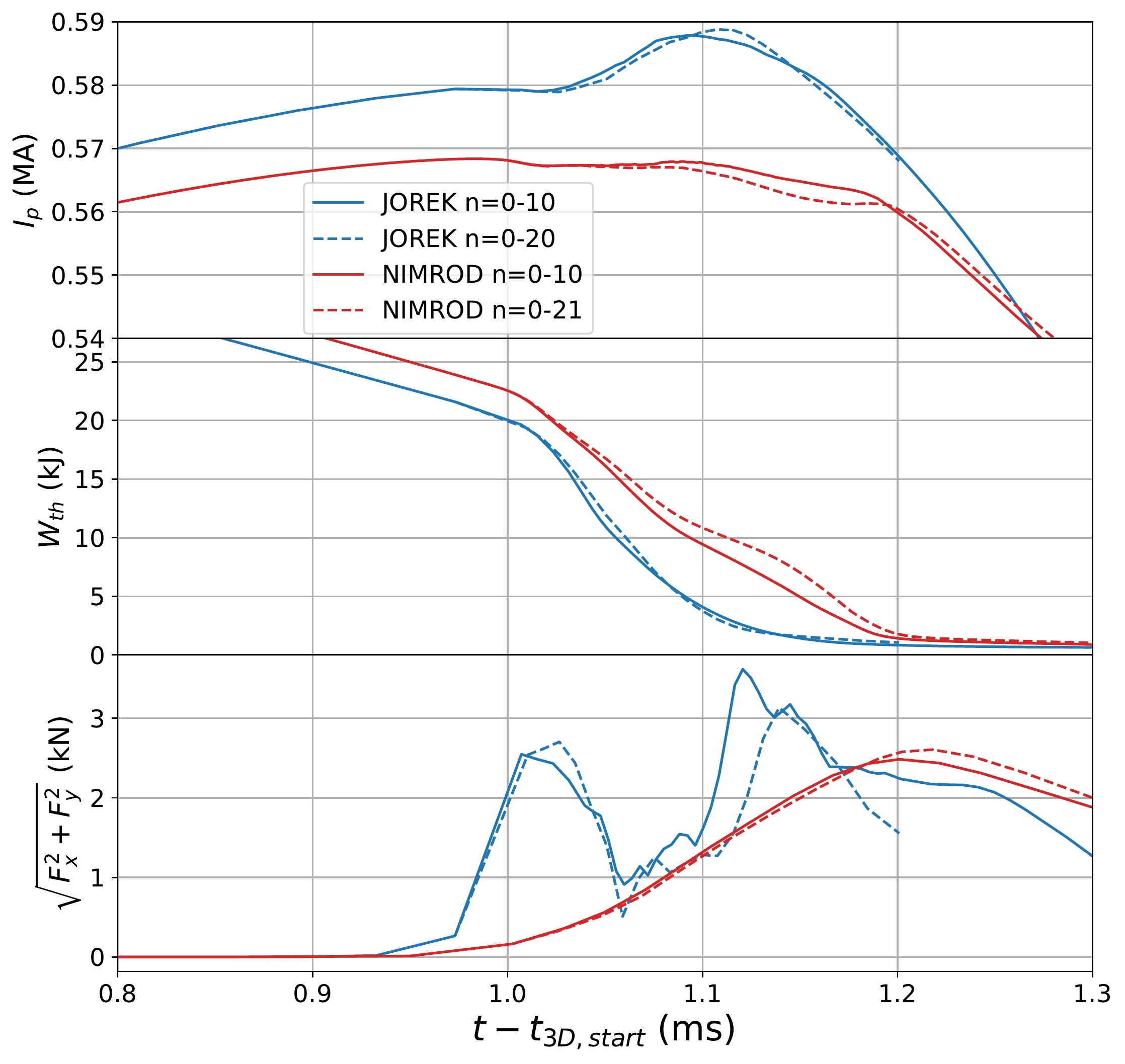}
\caption{Plasma current, thermal energy and horizontal force for two JOREK and NIMROD runs with different number of toroidal harmonics.}
\label{fig:res_scan}
\end{figure}

In NIMROD and in JOREK, the simulation was repeated with a larger number of toroidal harmonics, $n\in [0,20]$ instead of $n\in [0,10]$. The results are shown in figure \ref{fig:res_scan}, which indicate that high-n mode numbers do not change significantly the evolution of the important figures of merit (such as the plasma current, the thermal energy and the wall forces).

\subsection{Additional physics studies}
\label{sec:additional_physics}

\begin{figure}
\centering
  \includegraphics[width=0.95\textwidth]{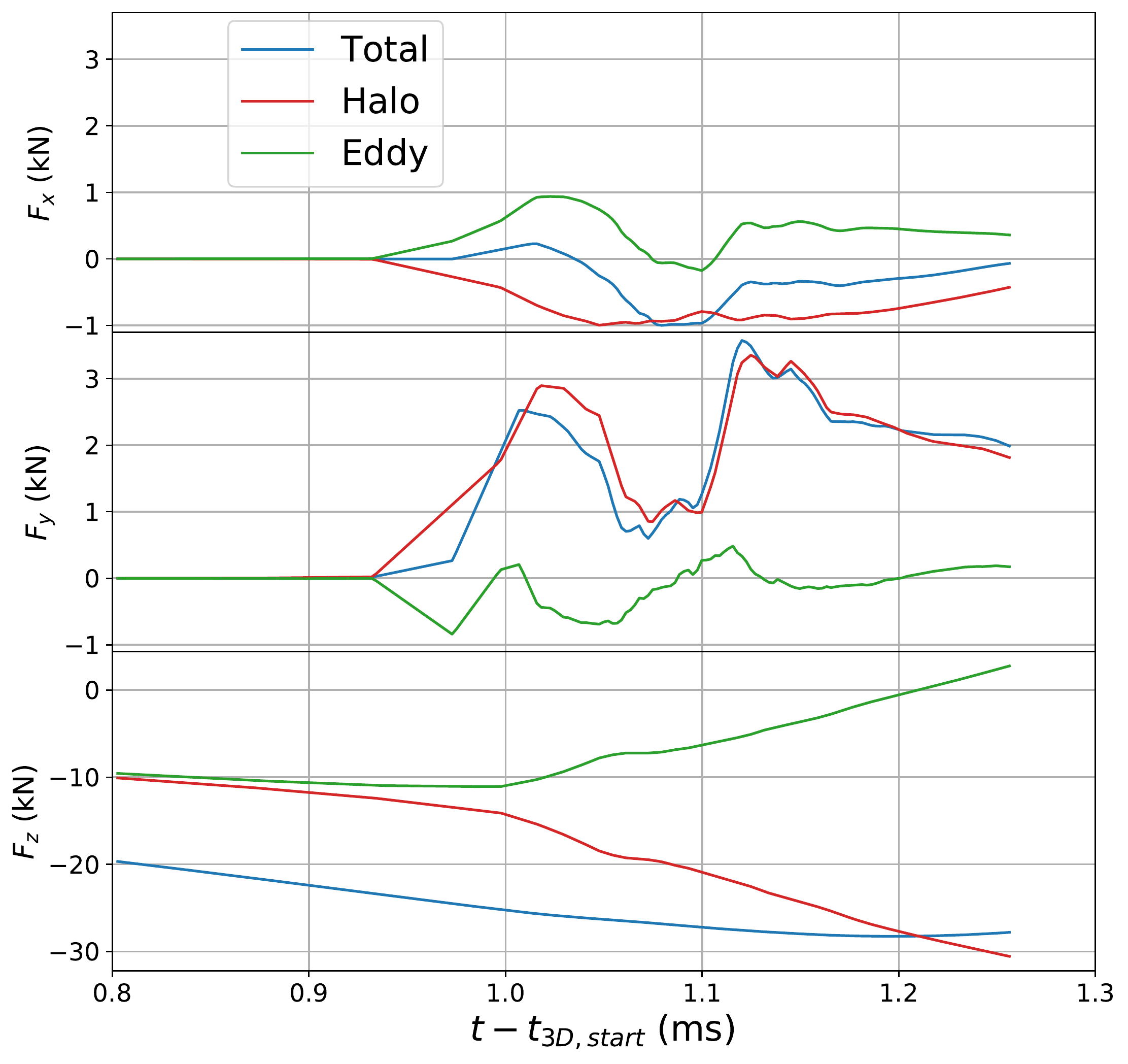}
\caption{Cartesian components of the total wall force in the JOREK simulation. The different legends denote the total wall force and the separate contributions of halo and eddy currents.}
\label{fig:force_origin}
\end{figure}

\subsubsection{Origin of the wall forces}
In this section, we study the origin of the wall forces with JOREK. As it was indicated in \ref{Ip_asym}, the boundary conditions used imply that for the reduced MHD model the plasma boundary acts as an ideal conductor in the poloidal direction. As a consequence, electric currents do not enter the wall ($\mathbf{J}\cdot\mathbf{n}|_{wall}=0$). In this sense, the plasma domain's boundary is not force-free. This feature allows to clearly separate between halo and eddy currents. Since only eddy currents flow on the JOREK-STARWALL wall, the wall force caused by eddy currents can be computed with
    
    \begin{equation}
        \mathbf{F}_{eddy}  =  \mathbf{F} -  \mathbf{F}_{halo} = \oint_{wall+} \mathbf{f} dS - \oint_{wall-} \mathbf{f} dS
    \end{equation}      
where $\mathbf{f}\equiv \left[(\mathbf{B}\cdot\mathbf{n})\mathbf{B} - \mathbf{n}B^2/2\right]/\mu_0 $ is the magnetic stress tensor projected in the normal direction to the wall. We have used the previous formula to calculate the different contributions to the total wall force as shown in figure \ref{fig:force_origin}. Although the eddy currents have a significant contribution, the horizontal force (mainly given by $F_y$) is governed by halo currents. Moreover, the eddy current force is in the opposite direction  with respect to the halo current force, and thus it reduces the total horizontal force. In the case of the vertical force, both halo and eddy currents have a similar contribution before the thermal quench.  After the thermal quench, the plasma resistivity is significantly increased  due to the drop in the plasma temperature. As a consequence, the core plasma current decays and it induces halo currents in the open field line region, which has a considerably large temperature due to the imposed B.C. $T_{e,edge}=14.6$ eV. As the induction of halo current and its associated wall force increases at a higher rate than the total wall force (which varies on the resistive wall time), the eddy current contribution to the total force must decrease as observed in \cite{Clauser_2019}. As a general conclusion for both forces, the JOREK simulation shows that the maximum forces are due to halo currents.

\subsubsection{Effect of parallel heat conductivity and viscosity}
\begin{figure}
\centering
  \includegraphics[width=0.7\textwidth]{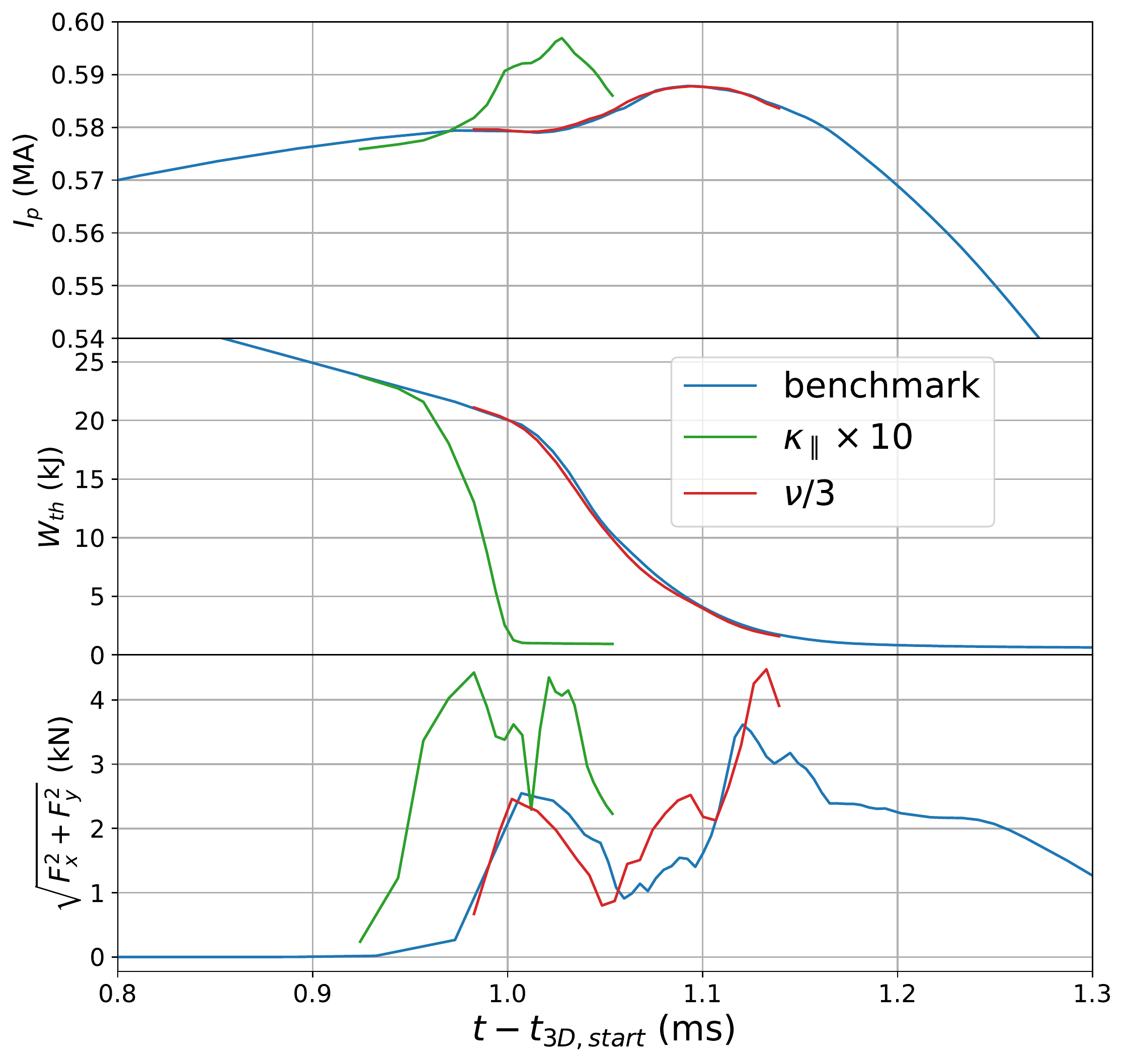}
\caption{Benchmark run with 10 times stronger parallel conductivity (green curves) and run with 3 times smaller viscosity (red curves) in JOREK. The baseline benchmark case (blue curves) is also plotted for reference. The quantities shown are the total toroidal current ($I_p$), the total thermal energy ($W_{th}$) and the total horizontal wall force.}
\label{fig:visco_kappa_effect}
\end{figure}

In this subsection a case with increased heat parallel conductivity and a case with decreased viscosity are presented. The results shown in figure \ref{fig:visco_kappa_effect} indicate that the viscosity plays a minor role for the evolution of $I_p$ and of the thermal energy. Although not large, reducing the viscosity produces perceptible effects on the wall force. Increasing the parallel conductivity produces a more dramatic effect on the results. The dynamics become faster and the loss of thermal energy takes  0.043 ms instead of 0.14 ms. Although it is expected that the increase of $\kappa_\parallel$ leads to a faster thermal quench (the heat is conducted faster along the open field lines), it is not clear that the thermal quench time has a linear dependence with the parallel heat conduction coefficient due to the effect of the thermal pressure on the field line topology. In this case when $\kappa_\parallel$ was increased by a factor of 10 the thermal quench time was reduced by a factor of 3. The magnitude of the horizontal force is significantly larger during the thermal quench although the maximum forces measured are similar (3.5-4.2 kN).

\section{Conclusions}
\label{sec:conclusions}

A benchmark case for a 3-dimensonal vertical displacement event has been presented based on an NSTX plasma. The case has been run with the MHD codes JOREK, M3D-C$^1$ and NIMROD for a 3-code comparison. The full run is divided into two phases: an axisymmetric run (2D) and a 3D run. Good agreement was found during the 2D phase for several figures of merit such as the plasma current, the thermal energy and the vertical position (as it was already checked in \cite{krebs2020axisymmetric}). The 3D phase was initiated when the plasma became limited by the wall instead of the lower X-point. 

\medskip

In spite of pronounced differences between physics models and numerical methods, a wide range of 3D features predicted by the three codes are in good agreement. For example, the three codes predict that the plasma becomes unstable to low-n external kink modes 0.85 - 1.1 ms after the plasma becomes limited by the wall. As indicated by JOREK, this happens when the $q$ value at the last closed flux surface falls below 2. The growth of these modes lead the stochastization of the magnetic field lines causing a loss of thermal energy on a time-scale of 0.14 ms in JOREK, 0.24 ms in M3D-C$^1$ and 0.20 ms in NIMROD. During the thermal quench similar filamentary structures can be observed in the pressure plots for the three codes (see figure \ref{fig:pres_evol}). These structures can be explained through chaotic magnetic field topology (see the Poincaré plots in figure \ref{fig:poincare}) and the fast parallel heat transport. The total vertical force onto the wall is in excellent agreement between the codes and the predicted maximum 3D horizontal forces are in the range of 2.5-3.5 kN. Moreover the horizontal force is only slowly rotating (less than one toroidal turn) after the thermal quench. Additional JOREK studies reveal that when the forces reach their maximum, they originate from  halo currents.

\medskip

The halo currents show mid-n filamentary patterns with large enough amplitude to reverse the sign of the normal current ($\mathbf{J}\cdot\mathbf{n}$) given by the $n=0$ component (see figure \ref{fig:Jnorm}). At the moment when the force reaches its maximum, the $n=1$ component becomes dominant. A scan in the number of Fourier harmonics used indicated that the wall force is sufficiently converged in toroidal resolution $n\in[0,10]$ to provide an adequate description of the modelled physics processes. Additional simulations show that the wall forces are weakly dependent on viscosity and that the choice of the parallel heat conductivity strongly influences the duration of the thermal quench but not the magnitude of the forces.

\medskip

Finally, important differences were observed for the toroidal asymmetry of $I_p$, showing that with similar boundary conditions for the normal velocity, the reduced MHD model of JOREK is not able to reproduce any $I_p$ asymmetries, which are of the order of a few per cent in NIMROD and M3D-C$^1$. This result motivates the implementation of a resistive wall in the full MHD model of JOREK. Nevertheless, the ansatz-based reduced MHD model is able to capture the 3D dynamics of the wall currents correctly, even for the large $\beta$ spherical tokamak plasma considered here.

\medskip

The consistent results among the three codes bring confidence for their use in disruption studies.  Moreover, important post-processing diagnostics were developed and validated during this work. Future efforts will focus on benchmarks of more complex simulations including Ohmic heating, radiation, realistic Spitzer resistivity, impurity injection, neutral particles and more advanced boundary conditions. Note that in order to have predictive capabilities and to validate the model predictions against the experiment, so that they can be applied with confidence to ITER, the latter effects need to be included (especially radiation and Ohmic heating) and therefore the present work is an important step towards realistic simulations.

\section*{Acknowledgements}
\label{sec:acknowledgements}
This work was supported by the ITER Monaco Fellowship Programme. The authors acknowledge access to the EUROfusion High Performance Computer (Marconi-Fusion) through EUROfusion funding to perform the presented simulations.

\medskip

ITER is the Nuclear Facility INB no. 174. This paper explores physics processes during the plasma operation of the tokamak when disruptions take place; nevertheless the nuclear operator is not constrained by the results presented here. The views and opinions expressed herein do not necessarily reflect those of the ITER Organization.

\medskip

This work was supported by US DOE Contracts No. DE-AC02-09CH11466 and No. DE-SC0018001, and the SciDAC Center for Tokamak Transient Simulations. Some of the computations presented in this report used resources of the National Energy Research Scientific Computing Center (NERSC), a U.S. Department of Energy Office of Science User Facility operated under Contract No. DE-AC02-05CH11231. The support from the EUROfusion Researcher Fellowship programme under the task agreement WP19-20-ERG-DIFFER/Krebs is gratefully acknowledged. Part of this work has been carried out within the framework of the EUROfusion Consortium and has received funding from the Euratom research and training programme 2014-2018 and 2019-2020 under grant agreement No 633053. The views and opinions expressed herein do not necessarily reflect those of the European Commission.

\medskip

The authors would like to thank Guido Huijsmans and Michael Lehnen for fruitful discussions about the contents of this article.

\bibliography{main}

\end{document}